# Fourier Domain Analysis performances of a *RESPER* probe

## Amplitude and Phase inaccuracies due to the *Round-Off* noise of *FFT* processors

Alessandro Settimi*

*Istituto Nazionale di Geofisica e Vulcanologia (INGV) –*
*Sezione Roma 2 -*
*via di Vigna Murata 605, I-00143 Rome, Italy*

*Corresponding author: Dr. Alessandro Settimi
Tel: +39-0651860719
Fax: +39-0651860397
Email: alessandro.settimi@ingv.it

## Index

## Introduction.

Electrical resistivity and relative dielectric permittivity are two independent physical properties which characterize the behaviour of bodies when these are excited by an electromagnetic field. The measurement of these properties provides crucial information regarding practical uses of bodies (for example, materials that conduct electricity) and for countless other purposes.

Some papers [Grard, 1990a,b][Grard and Tabbagh, 1991][Tabbagh et al., 1993][Vannaroni et al. 2004][Del Vento and Vannaroni, 2005] have proved that electrical resistivity and dielectric permittivity can be obtained by measuring complex impedance, using a system with four electrodes, but without requiring resistive contact between the electrodes and the investigated body. In this case, the current is made to circulate in the body by electric coupling, supplying the electrodes with an alternating electrical signal of *Low* or *Middle Frequency* (*LF-MF*). In this type of investigation, the range of optimal frequencies for resistivity values of the more common materials is between ≈*10kHz* and ≈*1MHz*. Once complex impedance has been acquired, the distributions of electrical resistivity and dielectric permittivity in the investigated body are estimated using well-known algorithms of inversion techniques.

Applying the same principle, but limited to the acquisition only of resistivity, there are various commercial instruments used in geology for investigating the first 2-5 meters underground both for the exploration of environmental areas and archaeological investigation [Samouëlian et al., 2005].

As regards the direct determination of the permittivity in subsoil, omitting geo-radar which provides an estimate by complex measurement procedures on radar-gram processing [Declerk, 1995][Sbartaï et al., 2006], the only technical instrument currently used is the so-called *Time-Domain Reflectometer* (*TDR*), which utilizes two electrodes inserted deep in the ground in order to acquire this parameter for further analysis [Mojid et al., 2003][Mojid and Cho, 2004].

## 1. *RESPER* probe.

Previous papers [Settimi et al., 2009-2010, a-c] presented a discussion of theoretical modelling and moved towards a practical implementation of a *RESPER* probe which acquires complex impedance in the field. A RESPER allows measurement of electrical RESistivity and dielectric PERmittivity using alternating current at LFs *(30kHz<f<300kHz)* or MFs *(300kHz<f<3MHz)*. By increasing the distance between the electrodes, it is possible to investigate the electrical properties of sub-surface structures to greater depth. In appropriate arrangements, measurements can be carried out with the electrodes slightly raised above the surface, enabling completely non-destructive analysis, although with greater error. The probe can perform immediate measurements on materials with high resistivity and permittivity, without subsequent stages of data analysis.

The paper [Settimi et al, 2009, b] has moved towards the practical implementation of electrical spectroscopy. In order to design a RESPER probe which measures the electrical resistivity and dielectric permittivity with inaccuracies below a prefixed limit (*10%*) in a band of LFs (*B=100kHz)*, the RESPER should be connected to an appropriate *Analogical to Digital Converter (ADC)*, which samples in uniform or in *Phase and Quadrature (IQ)* mode [Jankovic and Öhman, 2001]. If the probe is characterized by a galvanic contact with the surface, then the inaccuracies in the measurement of resistivity and permittivity, due to the uniform or IQ sampling ADC, can be analytically expressed. A large number of numerical simulations have proved that the performance depends on the selected sampler and that the IQ is preferable when compared to the uniform mode under the same operating conditions, i.e. number of bits and medium.

This report proposes to discuss the Fourier domain analysis performances of a RESPER probe. A uniform ADC, which is characterized by a sensible phase inaccuracy depending on frequency, is connected to a *Fast Fourier Transform (FFT)* processor, that is especially affected by a *round-off* amplitude noise linked to both the FFT register length and samples number. If the register length is equal to *32* bits, then the round-off noise is entirely negligible, else, once bits are reduced to *16*, a technique of compensation must occur. In fact, oversampling can be employed within a short time window, reaching a compromise between the needs of limiting the phase inaccuracy due to ADC and not raising too much the number of averaged FFT values sufficient to bound the round-off.

Finally, the appendix presents an outline of somewhat lengthy demonstrations needed to calculate the amplitude and especially phase inaccuracies due to the round-off noise of FFT processors.

## **2.** ***Analogical to Digital Converter (ADC).***

Typically, an ADC is an electronic device that converts an input analogical voltage (or current) to a digital number [Razavi, 1995]. A sampler has several sources of errors. Quantization error and (assuming the sampling is intended to be linear) non-linearity is intrinsic to any analogical-to-digital conversion. There is also a so-called *aperture error* which is due to *clock jitter* and is revealed when digitizing a time-variant signal (not a constant value). The accuracy is mainly limited by quantization error. However, a faithful reproduction is only possible if the sampling rate is higher than twice the highest frequency of the signal. This is essentially what is embodied in the Shannon-Nyquist sampling theorem.

There are currently a huge number of papers published in scientific literature, and the multifaceted nature of each one makes it difficult to present a complete overview of the ADC models available today. Technological progress, which is rapidly accelerating, makes this task even harder. Clearly, models of advanced digitizers must match the latest technological characteristics. Different users of sampler models are interested in different modelling details, and so numerous models are proposed in scientific literature: some of them describe specific error sources [Polge et al., 1975]; others are devised to connect conversion techniques and corresponding errors [Arpaia et al., 1999][Arpaia et al., 2003]; others again are devoted to measuring the effect of each error source in order to compensate it [Björsell and Händel, 2008]. Finally, many papers [Kuffel et al., 1991][Zhang and Ovaska, 1998] suggest general guidelines for different models.

In order to design a RESPER probe (Fig. 1.a) which measures the electrical conductivity $\sigma$ and the dielectric permittivity $\varepsilon_r$ of a subjacent medium with inaccuracies below a prefixed limit (*10%*) in a band of LFs ($B=100kHz$), the RESPER can be connected to an appropriate ADC, with bit resolution not exceeding *12*, thereby rendering the probe (voltage scale of *4V*) almost insensitive to the electric noise of the external environment ($\approx 1mV$) [Settimi et al., 2009-2010, a-c].

With the aim of investigating the physics of the measuring system, the inaccuracies in the complex impedance measured by the RESPER (Fig. 1.b) are provided.

In the stage downstream of the probe, the electrical voltage $V$ is amplified $V_V=A_V \cdot V$, then the intensity of current $I$ is transformed by a trans-resistance amplifier $V_I=A_R \cdot I$, and finally these signals are processed by the sampler. It follows that:

the inaccuracy $\Delta|Z|/|Z|$ for the amplitude of the complex impedance results from the negligible contributes $\Delta A_V/A_V$ and $\Delta A_R/A_R$, respectively for the voltage and trans-resistance amplifiers, and the predominant one $\Delta|V_V|/|V_V|$ for the amplitude of the voltage, due to the sampling,

$$\frac{\Delta|Z|}{|Z|}=\frac{\Delta A_V}{A_V}+\frac{\Delta A_R}{A_R}+2\frac{\Delta|V_V|}{|V_V|}\cong 2\frac{\Delta|V_V|}{|V_V|}, \qquad (2.1)$$

the inaccuracies for the amplitude of the voltage and the current intensity being equal, $\Delta|V_V|/|V_V| = \Delta|V_I|/|V_I|$;

instead, the inaccuracy $\Delta\Phi_Z/\Phi_Z$ for the initial phase of the complex impedance coincides with the one $\Delta\varphi_V/\varphi_V$ for the phase of the voltage, due to the sampler,

$$\frac{\Delta\Phi_Z}{\Phi_Z}=\frac{\Delta\varphi_V}{\varphi_V}, \qquad (2.2)$$

the initial phase of the current being null, $\varphi_I=0$.

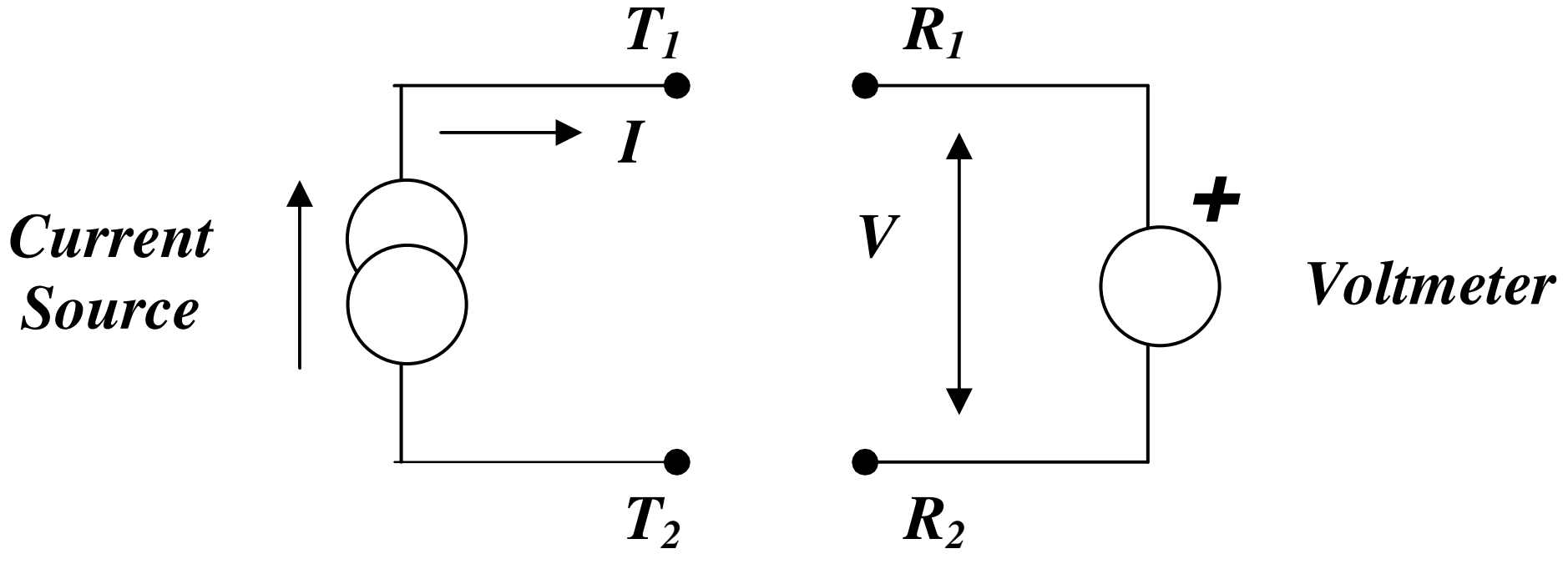

**Figure 1.a.** Equivalent circuit of a *RESPER* probe.

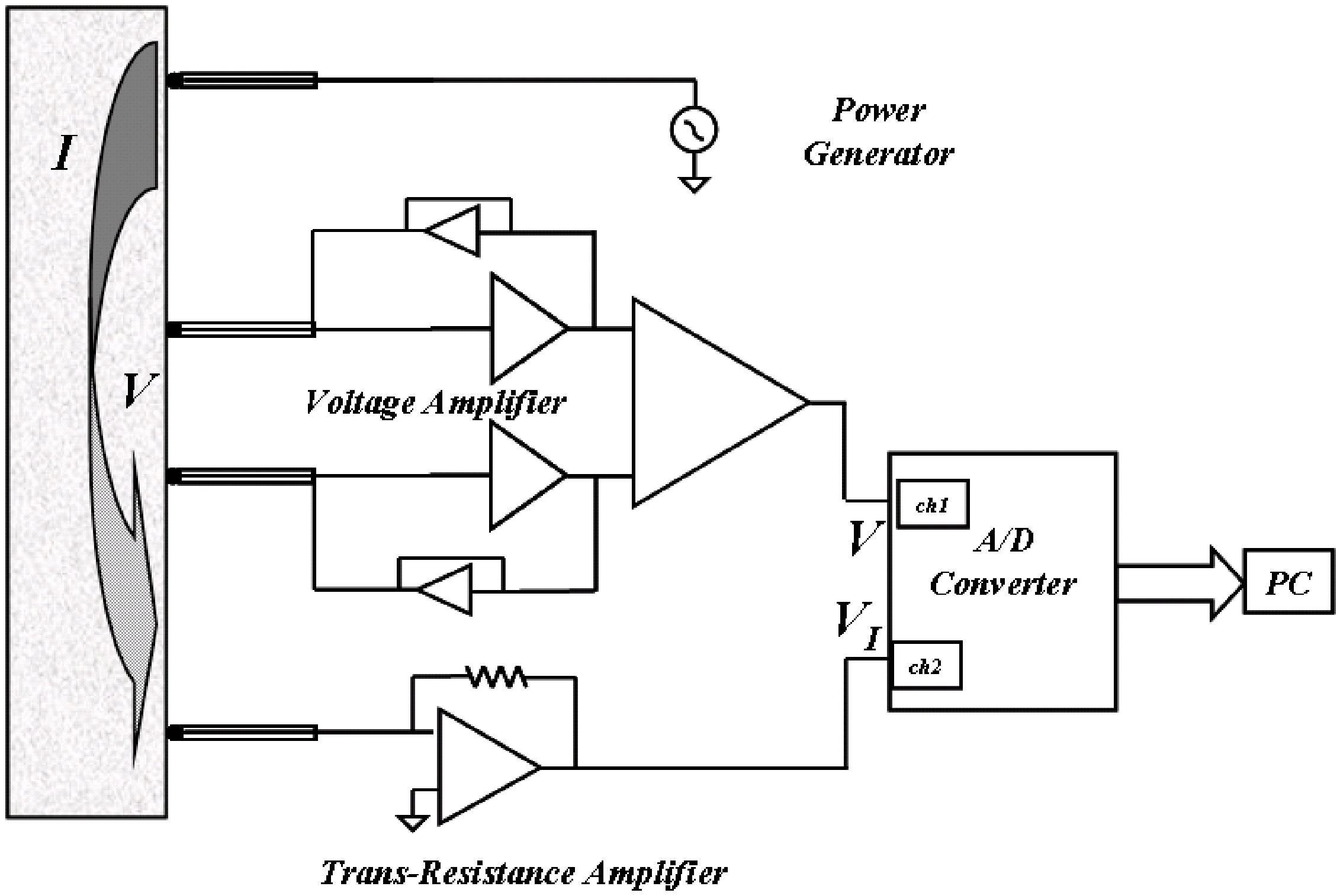

**Figure 1.b.** Block diagram of the measuring system, which is composed of: a series of four electrodes laid on the material to be investigated; an analogical circuit for the detection of signals connected to a high voltage sinusoidal generator; a digital acquisition system; and a personal computer. Starting from the left, the four electrodes can be seen laid on the block of material to be analyzed. Two electrodes are used to generate and measure the injected current (at a selected frequency), while the other two electrodes are used to measure the potential difference. In this way, two voltages are obtained: the first proportional to the current; the second proportional to the difference of potential. These voltages are digitized through an *Analogical to Digital converter (ADC)* connected to a personal computer for further processing. The real magnitudes hereby measured in the time domain are subsequently transformed into complex magnitudes in the frequency domain. From the ratio of the complex values, at the specific investigated frequency, it is possible to obtain the complex impedance. A program with an algorithm of numerical inversion allows the electrical resistivity and dielectric permittivity of the material to be obtained by measuring the complex impedance; in this way, the reliability of the measured data is immediately analyzed, proving very useful during a measurement program.

## 2.1. Uniform sampling ADC.

As concerns uniform sampling [Razavi, 1995], the inaccuracy $\Delta|Z|/|Z|_U(n)$ for $|Z|$ depends only on the bit resolution $n$, decreasing as the exponential function $2^{-n}$ of $n$ (Fig. 2.a),

$$\left.\frac{\Delta|Z|}{|Z|}\right|_U = \frac{1}{2^n}. \tag{2.3}$$

Instead, the inaccuracy $\Delta\Phi_Z/\Phi_Z|_U(f,f_S)$ for $\Phi_Z$ depends on both the working frequency $f$ of the RESPER and the rate sampling $f_S$ of the ADC, the inaccuracy being directly proportional to the frequency ratio $f/f_S$ (Fig. 2.b),

$$\left.\frac{\Delta\Phi_Z}{\Phi_Z}\right|_U = 2\frac{f}{f_S}. \tag{2.4}$$

As a consequence, for uniform sampling ADCs, the inaccuracy $\Delta\Phi_Z/\Phi_Z|_U(f,f_S)$ for the phase $\Phi_Z$ must be optimized in the upper frequency $f_{up}$, so when the probe performs measurements at the limit of its band $B$, i.e. $f_{up}=B$.

**(a)**

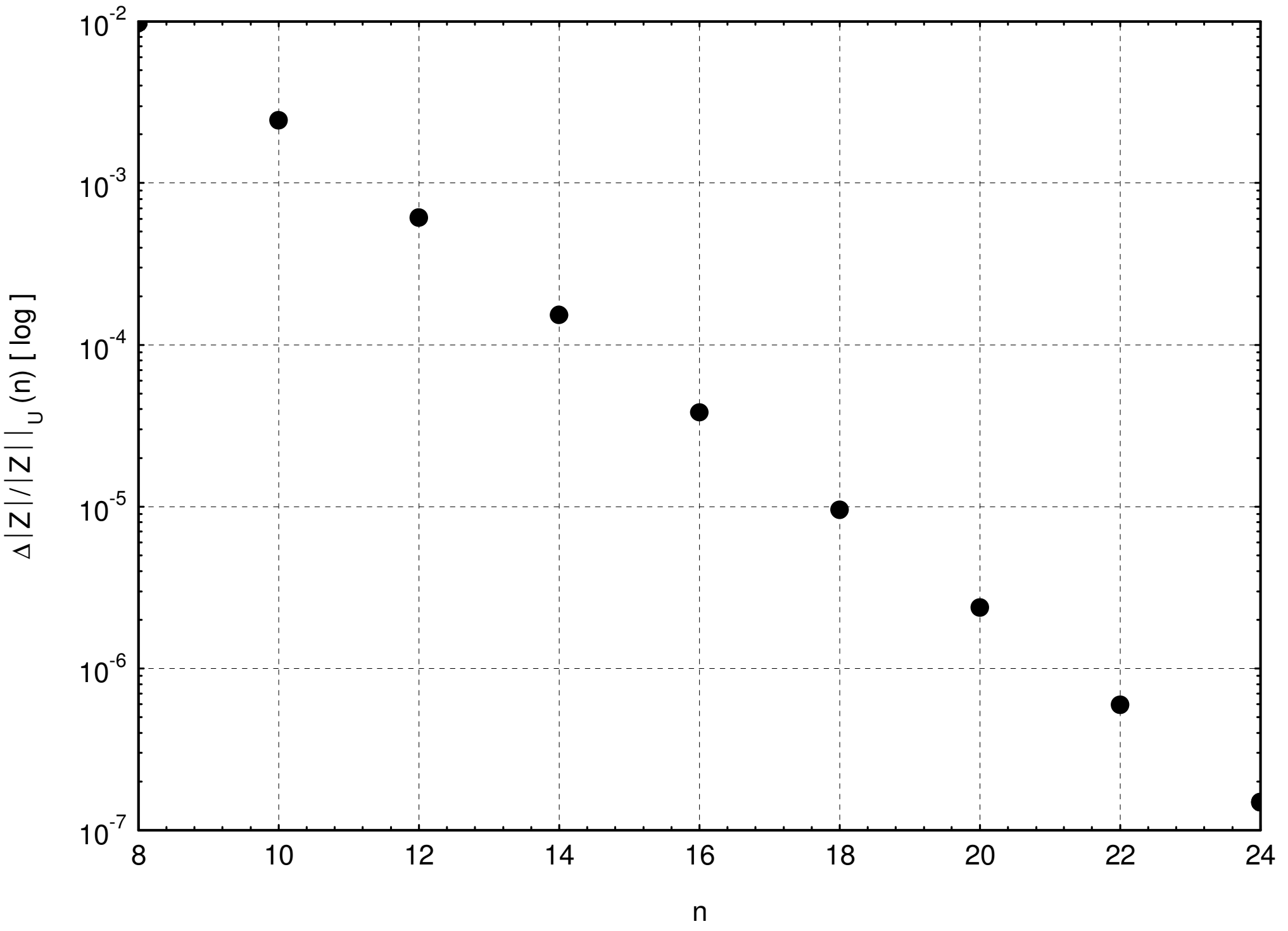

**(b)**

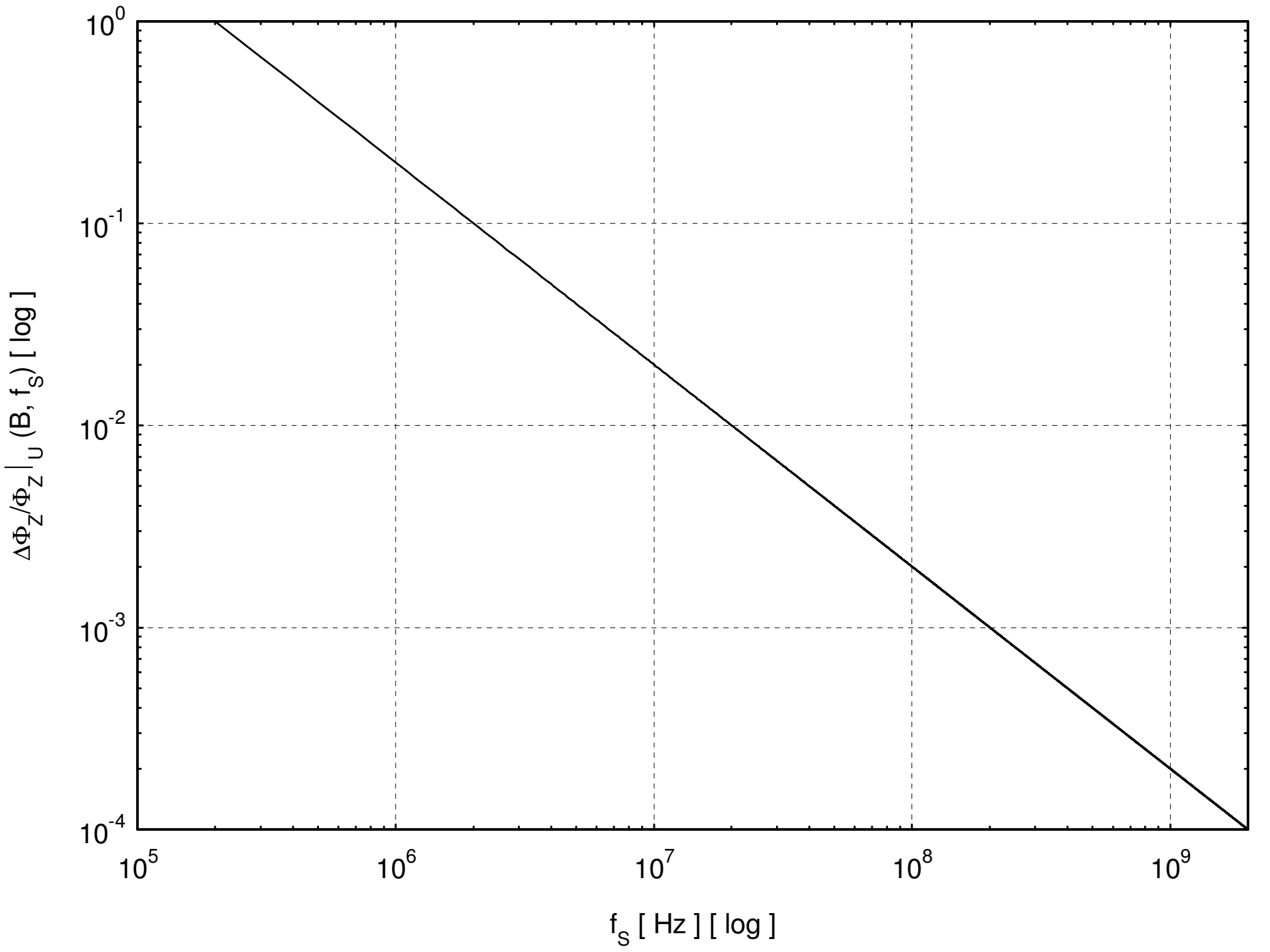

**Figure 2.** A class of uniform ADCs is specified by bit resolution $n$, ranging from *8 bit* to *24 bit*, and rate sampling $f_S$, in the band of frequency *[500 kHz, 2GHz]*: (a) semi-logarithmic plot for the inaccuracy $\Delta|Z|/|Z|_U(n)$ in the measurement of the amplitude for complex impedance, as a function of the resolution $n$; (b) Bode's diagram for the inaccuracy $\Delta\Phi_Z/\Phi_Z|_U(B,f_S)$ of the complex impedance in phase, plotted as a function of the rate $f_S$, when the RESPER works in the upper frequency at the limit of its band $B=100kHz$.

## 3. *Fast Fourier Transform (FFT)* processor and *round-off* noise.

In mathematics, the *Discrete Fourier Transform (DFT)* is a specific kind of Fourier transform, used in Fourier analysis. The DFT requires an input function that is discrete and whose non-zero values have a limited (finite) duration. Such inputs are often created by sampling a continuous function. Using the DFT implies that the finite segment which is analyzed is one period of an infinitely extended periodic signal; if this is not actually true, a window function has to be used to reduce the artefacts in the spectrum. In particular, the DFT is widely employed in signal processing and related fields to analyze the frequencies contained in a sampled signal. A key enabling factor for these applications is the fact that the DFT can be computed efficiently in practice using a *Fast Fourier Transform (FFT)* algorithm [Oppenheim et al., 1999].

It is important to understand the effects of finite register length in the computation. Specifically, arithmetic *round-off* is analyzed by means of a linear-noise model obtained by inserting an additive noise source at each point in the computation algorithm where round-off occurs. However, the effects of round-off noise are very similar among the different classes of FFT algorithms.

Generally, a FFT processor which computes $N$ samples, represented as $n_{FFT}+1$ bit signed fractions, is affected by a round-off noise which is added to the inaccuracy for complex impedance, in amplitude [Oppenheim et al., 1999][see Appendix]

$$\left.\frac{\Delta|Z|}{|Z|}\right|_{Round-off} = \frac{N}{2^{n_{FFT}-1}}, \tag{3.1}$$

and in phase [Dishan, 1995][Ming and Kang, 1996][see Appendix],

$$\left.\frac{\Delta\Phi_Z}{\Phi_Z}\right|_{Round-off} = \frac{1}{\pi\sqrt{N}\,2^{n_{FFT}}}. \tag{3.2}$$

So, maximizing the register length to $n_{FFT}=32$, the round-off noise is entirely negligible. Once that $n_{FFT}<32$, if the number of samples is increased $N>>1$, then the round-off noise due to FFT degrades the accuracy of complex impedance, so much more in amplitude (3.1) how much less in phase (3.2).

One can overcome this inconvenience by iterating the FFT processor for $A$ cycles, as the best estimate of one FFT value is the average of $A$ FFT repeated values. The improvement is that the inaccuracy for the averaged complex impedance, in amplitude and phase, consists of the error of quantization due to the uniform sampling ADC (2.3)-(2.4) and on the round-off noise due to FFT (3.1)-(3.2), the last term being decreased of $\sqrt{A}$, i.e.:

$$\frac{\Delta|Z|}{|Z|} = \left.\frac{\Delta|Z|}{|Z|}\right|_U + \frac{1}{\sqrt{A}}\left.\frac{\Delta|Z|}{|Z|}\right|_{Round-off} = \frac{1}{2^n} + \frac{1}{\sqrt{A}}\frac{N}{2^{n_{FFT}-1}}, \tag{3.3}$$

$$\frac{\Delta\Phi_Z}{\Phi_Z} = \left.\frac{\Delta\Phi_Z}{\Phi_Z}\right|_U + \frac{1}{\sqrt{A}}\left.\frac{\Delta\Phi_Z}{\Phi_Z}\right|_{Round-off} = 2\frac{f}{f_S} + \frac{1}{\sqrt{A}}\frac{1}{\pi\sqrt{N}\,2^{n_{FFT}}}. \tag{3.4}$$

Once reduced the register length to $n_{FFT}\leq 16$, only if the FFT processor performs the averages during a number of cycles

$$\left(\frac{N}{2^{n_{FFT}-n-1}}\right)^2 << A \leq N^2, \tag{3.5}$$

then the round-off noise due to FFT can be neglected with respect to the quantization error due to uniform ADC, in amplitude

$$\frac{\Delta|Z|}{|Z|} \cong \frac{1}{2^n} + \frac{1}{2^{n_{FFT}-1}} \approx \left.\frac{\Delta|Z|}{|Z|}\right|_U = \frac{1}{2^n}, \tag{3.6}$$

and especially in phase

$$\frac{\Delta\Phi_Z}{\Phi_Z} \cong 2\frac{f}{f_S} + \frac{1}{\pi N\sqrt{N}\,2^{n_{FFT}}} \approx \left.\frac{\Delta\Phi_Z}{\Phi_Z}\right|_U = 2\frac{f}{f_S}. \tag{3.7}$$

So, the round-off noise due to FFT is compensated. The quantization error due to ADC decides the accuracy for complex impedance: it is constant in amplitude, once fixed the bit resolution $n$, and can be limited in phase, by an oversampling technique $f_S >> f$.

In the limit of the Shannon-Nyquist theorem, an electric signal with band of frequency $B$ must be sampled at the minimal rate $f_S = 2B$, holding $N_{min}$ samples in a window of time $T$. Instead, in the hypothesis of oversampling, the signal can be sampled holding the same number of samples $N_{min}$ but in a shorter time window $T/R_O$, due to an high ratio of sampling:

$$R_O = \frac{f_S}{2B} >> 1. \tag{3.8}$$

This is equivalent to the operating condition such that, during the time window,

$$T = \frac{N_{\min}}{2B}, \tag{3.9}$$

the uniform over-sampling ADC holds a samples number

$$N = R_O \cdot N_{\min} >> N_{\min} \tag{3.10}$$

which is linked to the number of cycles iterated by the FFT processor:

$$A \cong N^2 = R_O^2 \cdot N_{\min}^2 >> 1. \tag{3.11}$$

As comments on eqs. (3.9)-(3.11), a low number of samples $N_{min}$, corresponding to the Shannon-Nyquist limit, shortens the time window (3.9). An high oversampling ratio lowers the phase inaccuracy although it raises the samples number hold by uniform ADC and especially the cycles number iterated by FFT; however, even a minimal oversampling ratio $R_{O,min}$ limits the phase inaccuracy with the advantage of not raising too much the samples number hold by ADC (3.10) and especially the cycles number iterated by FFT (3.11).

A RESPER probe (frequency band $B$) shows a galvanic contact with the subjacent non-saturated medium (terrestrial soil or concrete with low dielectric permittivity, $\varepsilon_r = 4$, and high electrical resistivity, $1/\sigma_S = 3 \cdot 10^3\ \Omega \cdot m$, $1/\sigma_C = 1 \cdot 10^4\ \Omega \cdot m$). It is required that the inaccuracy $\Delta\varepsilon_r/\varepsilon_r(f, f_S, n)$ in the measurement of permittivity $\varepsilon_r$ is below a prefixed limit $\Delta\varepsilon_r/\varepsilon_r|_{fixed}$ $(10\% \div 15\%)$ within the band $B$ $(100kHz)$. As a first result, if the samples number satisfying the Shannon-Nyquist theorem is minimized, i.e. $N_{min}=2$, then the time window for sampling is shortened to $T = N_{min}/(2B) = 1/B \approx 10\mu s$. In order to analyze the complex impedance measured by the RESPER in Fourier domain, a uniform ADC can be connected to a FFT processor, being affected by a round-off amplitude noise. As a conclusive result, a technique of compensation must occur. The ADC must be specified by: a minimal bit resolution $n \leq 12$, thereby rendering the probe almost insensitive to the electric noise of the external environment; and a minimal over-sampling rate $f_S$, which limits the ratio $R_O = f_S/(2B)$, so the actual samples number $N = R_O \cdot N_{min}$ is up to one hundred (soil, $f_S = 10MHz$, $R_O = 50$, $N \approx 100$)(concrete, $f_S = 5MHz$, $R_O = 25$, $N \approx 50$). Moreover, even if the FFT register length is equal to $n_{FFT} = 16$, anyway the minimal rate $f_S$ ensures a number of averaged FFT values $A \leq N^2$ even up to ten thousand, necessary to bound the round-off noise (soil, $A \approx 10^4$)(concrete, $A \approx 2.5 \cdot 10^3$) (Fig. 3)(Tab. 1) [Settimi et al., 2009-2010, a-c].

**(a)**

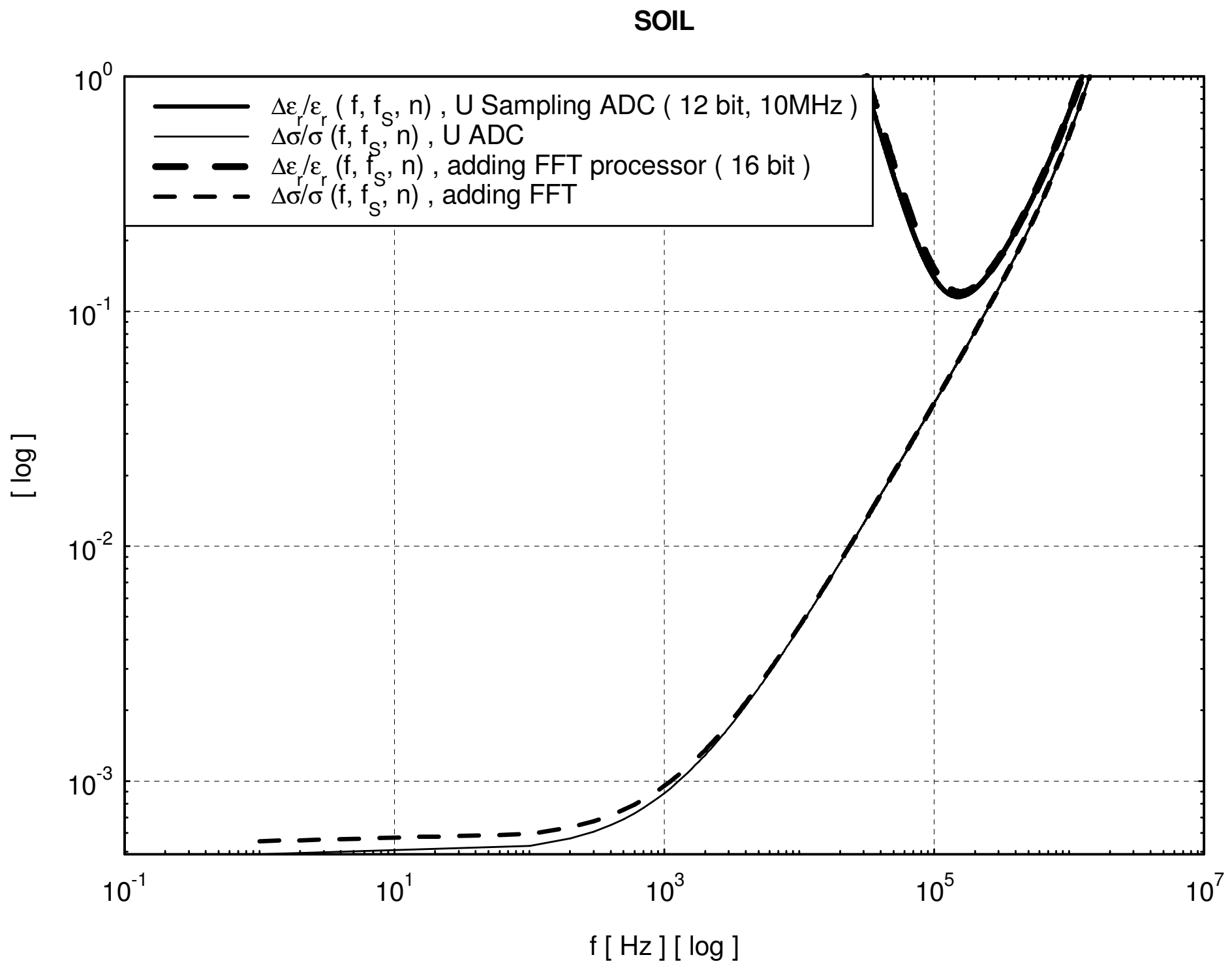

**(b)**

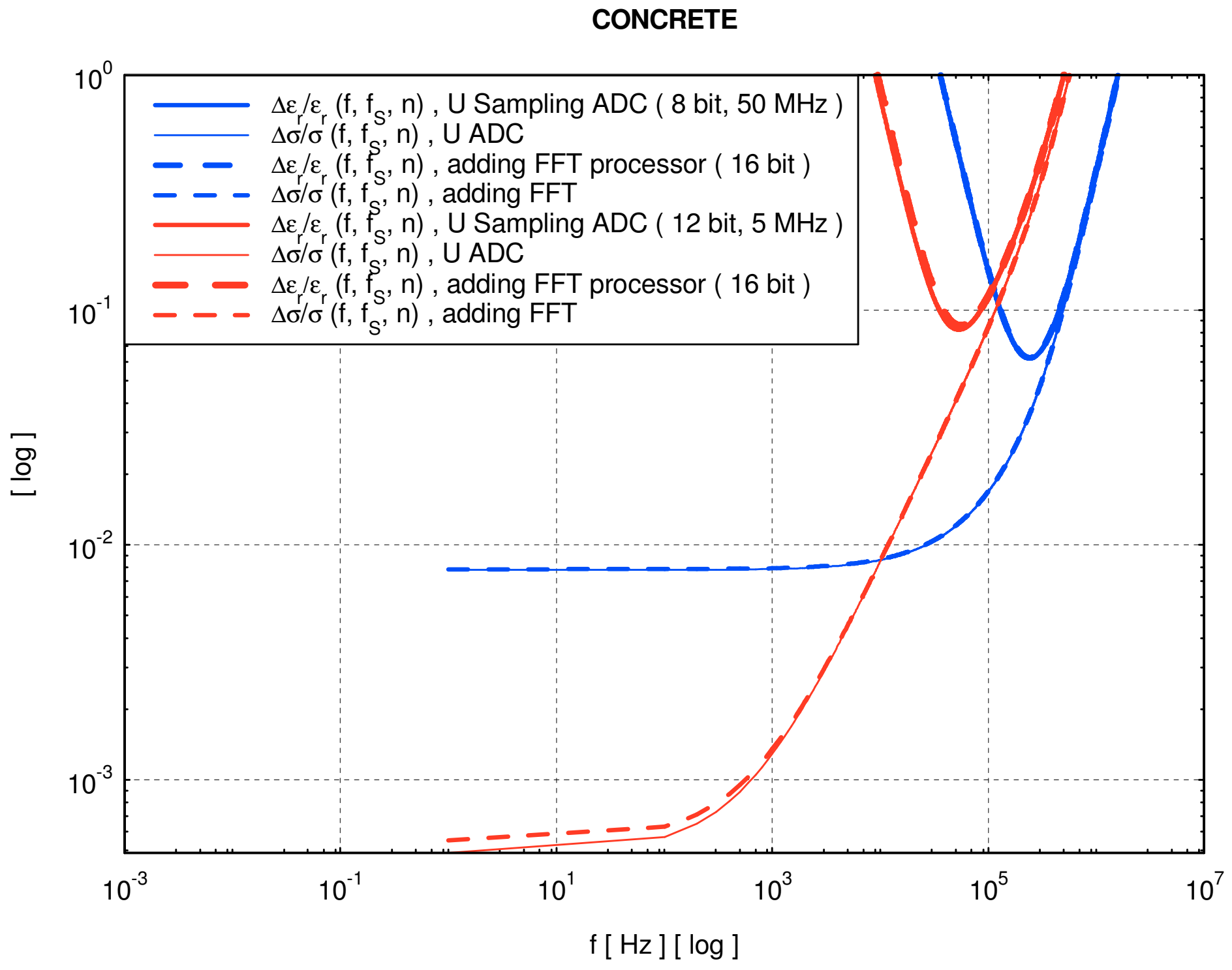

**Figure 3.** The probe is connected to an uniform ADC of minimal bit resolution $n \leq 12$ and over-sampling rate $f_S$, (*12 bit, 10 MHz*)(a) and (*8 bit, 50 MHz*) or (*12 bit, 5 MHz*)(b), in addition to a *Fast Fourier Transform (FFT)* processor of register length $n_{FFT} = 16$ which ensures inaccuracies $\Delta\varepsilon_r/\varepsilon_r(f)$ and $\Delta\sigma/\sigma(f)$ in the measurement of permittivity $\varepsilon_r$ and conductivity $\sigma$ below a prefixed limit, *15%* referring to (a) and *10%* for (b), within the frequency band $B=100kHz$ (Tab. 1) [Settimi et al., 2009-2010, a-c].

**(a)**

| *Soil* ($\varepsilon_r = 4$, $\rho = 3000\ \Omega\cdot m$) | **U Sampling ADC (n = 12, $f_S$ = 10 MHz) + FFT ($n_{FFT}$ = 16)** |
|---|---|
| $T = N_{min}/2B$ | ≈ 10 μs |
| $N = N_{min}\cdot(f_S/2B)$ | ≈ 100 ($2^7 = 128$) |
| $A \le N^2$ | ≈ $10^4$ |
| $f_{opt}, f_{min}, f_{max}$ ($\Delta\varepsilon_r/\varepsilon_r, \Delta\sigma/\sigma \le 0.15$) | 156.256 kHz, 99.68 kHz, 261.559 kHz |

**(b)**

| *Concrete* ($\varepsilon_r = 4$, $\rho = 10000\ \Omega\cdot m$) | **U Sampling ADC (n = 8, $f_S$ = 50 MHz) + FFT ($n_{FFT}$ = 16)** | **U Sampling ADC (n = 12, $f_S$ = 5 MHz) + FFT ($n_{FFT}$ = 16)** |
|---|---|---|
| $T = N_{min}/2B$ | ≈ 10 μs | |
| $N = N_{min}\cdot(f_S/2B)$ | ≈ 500 ($2^9 = 512$) | ≈ 50 ($2^6 = 64$) |
| $A \le N^2$ | ≈ $2.5\cdot10^5$ | ≈ $2.5\cdot10^3$ |
| $f_{opt}, f_{min}, f_{max}$ | 241.906 kHz, 99.007kHz, 591.411 kHz ($\Delta\varepsilon_r/\varepsilon_r, \Delta\sigma/\sigma \le 0.15$) | 55.344 kHz, 38.195kHz, 83.642 kHz ($\Delta\varepsilon_r/\varepsilon_r, \Delta\sigma/\sigma \le 0.1$) |

**Table 1.** Refer to the caption of fig. 3. A RESPER probe is connected to an uniform ADC (Shannon-Nyquist theorem: limit of samples number, $N_{min} = 2$), in addition to a FFT processor with *round-off* noise ($T$, time window; $N$, actual samples number; $A$, cycles number averaging FFT values). Optimal, minimum and maximum working frequencies, $f_{opt}$, $f_{min}$ and $f_{max}$, for measurements performed on terrestrial soil (a) and concrete (b) [Settimi et al., 2009-2010, a-c].

## Appendix: round-off noise of a FFT processor.

The *Discrete Fourier Transform (DFT)* plays an important role in the analysis, design, and implementation of discrete-time signal-processing algorithms and systems [Oppenheim et al., 1999]. The basic properties of the Fourier transform and discrete Fourier transform make it particularly convenient to analyze and design systems in the Fourier domain. Equally important is the fact that efficient algorithms exist for explicitly computing the DTF. As a result, the DTF is an important component in many practical applications of discrete-time systems.

As discussed in Oppenheim (1999), the DTF is identical to samples of the Fourier transform at equally spaced frequencies. Consequently, computation of the $N$-point DTF corresponds to the computation of the $N$ samples of the Fourier transform at $N$ equally spaced frequencies, $\omega_k=2\pi k/N$, i.e. at $N$ points on the unit circle in the complex plane. Oppenheim considers techniques for computation of the discrete Fourier transform. The periodicity and symmetry of the complex factor $W_N^{kn} = e^{-j(2\pi/N)kn}$ can be exploited to increase the efficiency of DFT computations. However, the major emphasis is on *Fast Fourier Transform (FFT)* algorithms. The *decimation-in-time* and *decimation-in-frequency* classes of FFT algorithms are described in some detail, and even some of the implementation considerations, such as indexing and coefficient quantization. Much of the detailed discussion concerns algorithms that require $N$ to be a power of 2, since these algorithms are easy to understand, simple to program, and most often used.

Oppenheim (1999) has discussed effects of finite word length in DFT computations. Linear-noise models are used to show that the *Noise-to-Signal Ratio* of a DFT computation varies differently with the length of the sequence, depending on how scaling is done. Oppenheim also comments briefly on the use of floating-point representations.

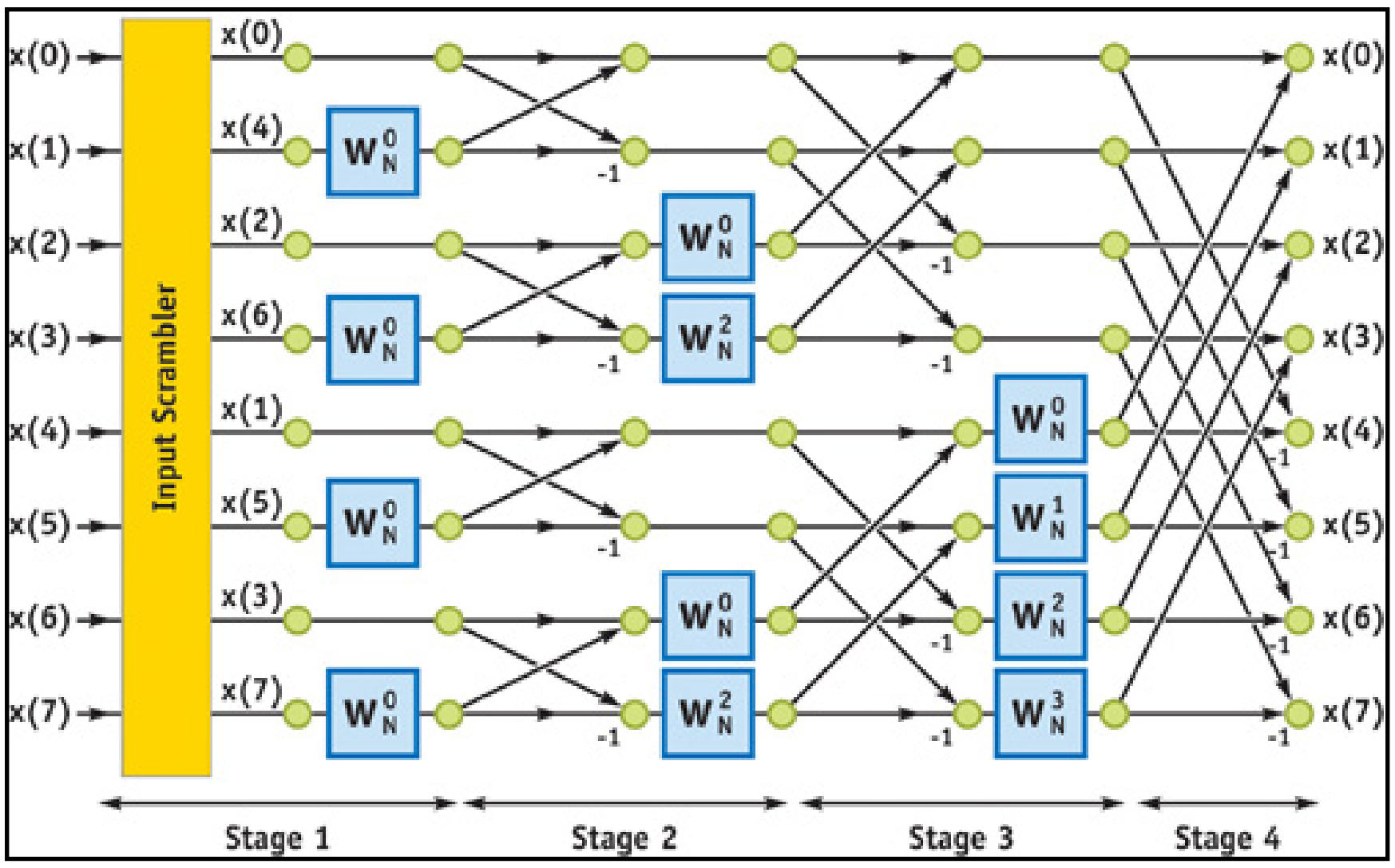

**Figure 4**. Flow graph for decimation-in-time FFT algorithm.

**Amplitude *Noise-to-Signal Ratio*.**

A flow graph depicting the decimation-in-time algorithm for $N=8$ is shown in fig. 4. Some key aspects of this diagram are common to all standard radix-2 algorithms. The DFT is computed in $v=log_2N$ stages. At each stage a new array of $N$ numbers is formed from the previous array by linear combinations of the elements, taken two at a time. The ν-th array contains the desired DFT.

The round-off noise is modelled by associating an additive noise generator with each fixed-point multiplication [Oppenheim et al., 1999].

Since, in general, the input to the FFT is a complex sequence, each of the multiplications is complex and thus consists of four real multiplications. Assume that the errors due to each real multiplication have the following properties:

1. The errors are uniformly distributed random variables over the range $-(1/2)\cdot 2^{n_{FFT}}$ to $+(1/2)\cdot 2^{n_{FFT}}$, where numbers are represented as $(n_{FFT}+1)$-bit signed fractions. Therefore, each error source has variance $2^{-2n_{FFT}}/12$.
2. The errors are uncorrelated with one another.
3. all the errors are uncorrelated with the input and, consequently, also with the output.

Since each of the four noise sequences is uncorrelated zero-mean white noise and all have the same variance,

$$\sigma^2_{n_{FFT}} = 4\cdot\frac{2^{-2n_{FFT}}}{12} = \frac{1}{3}\cdot 2^{-2n_{FFT}} . \tag{A.1}$$

To determine the mean-square value of the output noise at any output node, one must account for the contribution from each of the noise sources which propagate to that node.

The following observations can be made from the flow graph of fig. 4:

1. The transmission function from any node in the flow graph to any other node to which it is connected is multiplicated by a complex constant of unit magnitude (because each branch transmittance is either unity or an integer power of $W_N$).
2. Each output node connects to seven butterflies in the flow graph. In general, each output node would connect to $(N\text{-}1)$ butterflies.

These observations can be generalized to the case of $N$ an arbitrary power of $2$.

With these assumptions, then, the mean square value of the output noise in the $k$-th DFT value, $F[k]$, is given by [Oppenheim et al., 1999]

$$\left\langle \left|F\left[k\right]\right|^2 \right\rangle = (N-1)\sigma^2_{n_{FFT}} , \tag{A.2}$$

which, for large $N$, can be approximated as:

$$\left\langle \left|F\left[k\right]\right|^2 \right\rangle \cong N\sigma^2_{n_{FFT}} . \tag{A.3}$$

According to this result, the mean square value of the output noise is proportional to $N$, the number of points transformed. The effect of doubling $N$, or adding another stage in the FFT, is to double the mean-square value of the output noise. Note that for FFT algorithms, a double-length accumulator does not help to reduce round-off noise, since the outputs of the butterfly computation must be stored in $(n_{FFT}+1)$-bit registers at the output of each stage.

In implementing an FFT algorithm with fixed-point arithmetic, one must ensure against overflow. If the magnitude of the output of the FFT is less than unity, then the magnitude of the points in each array must be less than unity, i.e. there will be no overflow in any of the arrays.

To express this constraint as a bound on the input sequence, note that the condition

$$\left|x\left[n\right]\right| < \frac{1}{N} \quad , \quad 0 \le n \le N-1 , \tag{A.4}$$

is both necessary and sufficient to guarantee that

$$\left|X\left[k\right]\right| < 1 \quad , \quad 0 \le k \le N-1 . \tag{A.5}$$

This follows from the definition of the DFT, since:

$$\left|X[k]\right| = \left|\sum_{n=0}^{N-1} x[n] W_N^{kn}\right| \le \left|\sum_{n=0}^{N-1} x[n]\right| \quad , \quad k = 0, 1, \ldots N-1. \tag{A.6}$$

Then eq. (A.4) is sufficient to guarantee that there will be no overflow for all stages of algorithm.

To obtain an explicit expression for the Noise-to-Signal Ratio at the output of the FFT algorithm, consider an input in which successive sequence values are uncorrelated, i.e. a white-noise input signal. Also, assume that the real an imaginary parts of the input sequence are uncorrelated and that each has an amplitude density which is uniform between $-1/(\sqrt{2}\,N)$ and $+1/(\sqrt{2}\,N)$ [Note that this signal satisfies eq. (A.4)]. Then the average squared magnitude of the complex input sequence is:

$$\sigma_x^2 = \left\langle \left|x[n]\right|^2 \right\rangle = \frac{1}{3N^2}. \tag{A.7}$$

The DTF of the input sequence is

$$X[k] = \sum_{n=0}^{N-1} x[n] W_N^{kn}, \tag{A.8}$$

from which it can be shown that, under the foregoing assumptions on the input,

$$\left\langle \left|X[k]\right|^2 \right\rangle = \sum_{n=0}^{N-1} \left\langle \left|x[n]\right|^2 \right\rangle \left|W_N^{kn}\right|^2 = N\sigma_x^2 = \frac{1}{3N}. \tag{A.9}$$

Combining eqs. (A.3) and (A.9), it is obtained [see eq. (3.1)]:

$$\frac{\left\langle \left|F[k]\right|^2 \right\rangle}{\left\langle \left|X[k]\right|^2 \right\rangle} \cong 3N^2\sigma_{n_{FFT}}^2 = N^2 2^{-2n_{FFT}}. \tag{A.10}$$

According to eq. (A.10), the noise-to-signal ratio increases as $N^2$, or *1* bit per stage. That is, if $N$ is doubled, corresponding to adding one additional stage to the FFT, then to maintain the same noise-to-signal ratio, *1* bit must be added to the register length. The assumption of a white-noise input signal is, in fact, not critical here. For a variety of other inputs, the noise-to-signal ratio is still proportional to $N^2$, with only the constant of proportionality changing.

The preceding analysis shows that scaling to avoid overflow is the dominant factor in determining the noise-to-signal ratio of fixed-point implementations of FFT algorithms. Therefore, floating-point arithmetic should improve the performances of these algorithms. The effect of floating point round-off on the FFT is analyzed both theoretically and experimentally by Gentleman and Sande (1966), Weinstein (1969), and Kaneko and Liu (1970) (see references therein [Oppenheim et al., 1999]). The investigations show that, since scaling is no longer necessary, the decrease of noise-to-signal ratio with increasing $N$ is much less dramatic than for fixed-point arithmetic.

For example, Weinstein (1969) showed theoretically that the noise-to-signal ratio is proportional to $v$ for $N=2^v$, rather than proportional to $N$ as in the fixed-point case. Therefore, quadrupling $v$ (raising $N$ to the fourth power) increases the noise-to-signal ratio by only *1* bit.

**Phase Noise-to-Signal Ratio.**

Firstly, let us consider the *k*-th value of the DTF

$$X[k]=\sum_{n=0}^{N-1}x[n]W_N^{kn}=\sum_{n=0}^{N-1}x[n]e^{-j(2\pi/N)kn} \quad , \quad k=0,1,\dots N-1, \qquad (A.11)$$

corresponding to a complex sequence:

$$x[n]=x_R[n]+jx_I[n]=|x[n]|\,\mathrm{e}^{j\varphi[n]} \quad , \quad n=0,1,\dots N-1. \qquad (A.12)$$

The phase $\Phi[k]$ of the *k*-th DFT value $X[k]=|X[k]|e^{j\Phi[k]}$ can be calculated as

$$\Phi[k]=\operatorname{Im}\ln X[k], \qquad (A.13)$$

instead the phase $\varphi[n]-(2\pi/N)kn$ of the *n*-th sequence term $x[n]e^{-j(2\pi/N)kn}=|x[n]|e^{j\varphi[n]}e^{-j(2\pi/N)kn}$ as:

$$\varphi[n]-\frac{2\pi}{N}kn=\operatorname{Im}\ln\left\{x[n]e^{-j(2\pi/N)kn}\right\}. \qquad (A.14)$$

Applying a special inequality

$$\ln\sum_{n=0}^{N-1}x[n]e^{-j(2\pi/N)kn}\le\frac{1}{N}\sum_{n=0}^{N-1}\ln\left\{x[n]e^{-j(2\pi/N)kn}\right\}, \qquad (A.15)$$

the phase $\Phi[k]$ of the *k*-th DFT value (A.11) can be superiorly limited by the sum of phases $\varphi[n]$ for the complex sequence (A.12):

$$\begin{aligned}\Phi[k]&=\operatorname{Im}\ln X[k]=\operatorname{Im}\ln\sum_{n=0}^{N-1}x[n]e^{-j(2\pi/N)kn}\le\\ &\overset{eq.(A.15)}{\le}\frac{1}{N}\sum_{n=0}^{N-1}\operatorname{Im}\ln\left\{x[n]e^{-j(2\pi/N)kn}\right\}=\frac{1}{N}\sum_{n=0}^{N-1}\left\{\varphi[n]-\frac{2\pi}{N}kn\right\}=\frac{1}{N}\left\{\sum_{n=0}^{N-1}\varphi[n]-\frac{2\pi}{N}k\sum_{n=0}^{N-1}n\right\}= \quad .\\ &\overset{eq.(A.35)}{=}\frac{1}{N}\left\{\sum_{n=0}^{N-1}\varphi[n]-\frac{2\pi}{N}k\frac{N(N-1)}{2}\right\}=\frac{1}{N}\left\{\sum_{n=0}^{N-1}\varphi[n]-\pi k(N-1)\right\}\end{aligned} \qquad (A.16)$$

It follows that: the mean value $\langle\Phi[k]\rangle$ of the *k*-th DFT value phase can be expressed as a linear combination of all the mean values $\langle\varphi[n]\rangle$ for the sequence phases and the mean value $\langle k\rangle$ of the index $k=0,1\dots N-1$,

$$\langle\Phi[k]\rangle\cong\frac{1}{N}\left\{\sum_{n=0}^{N-1}\langle\varphi[n]\rangle-\pi\langle k\rangle(N-1)\right\}; \qquad (A.17)$$

then, if the variables $\varphi[n]$ and $k$ are uncorrelated, the variance $\sigma_\Phi^2$ is a combination of both the variances $\sigma_\varphi^2$ and $\sigma_k^2$,

$$\sigma_\Phi^2\cong\frac{1}{N^2}\left\{\sum_{n=0}^{N-1}\sigma_{\varphi[n]}^2+\pi^2\sigma_k^2(N-1)^2\right\}=\frac{1}{N^2}[N\sigma_\varphi^2+\pi^2\sigma_k^2(N-1)^2]; \qquad (A.18)$$

finally, as $\sigma_\Phi^2=\langle\Phi^2[k]\rangle-\langle\Phi[k]\rangle^2$, the mean square value:

$$\langle\Phi^2[k]\rangle=\sigma_\Phi^2+\langle\Phi[k]\rangle^2. \qquad (A.19)$$

Secondly, consider a stochastic complex sequence $x[n]$ consisting of $N$ values [see eq. (A.12)], whose imaginary and real parts are uniform random variables between $-1/(\sqrt{2}\,N)$ and $+1/(\sqrt{2}\,N)$, defined by a density of probability

$$p_x(x)=\frac{N}{\sqrt{2}}rect_{1/\sqrt{2}N}(x)=\begin{cases}\dfrac{N}{\sqrt{2}} & , \; -\dfrac{1}{\sqrt{2}N}<x<\dfrac{1}{\sqrt{2}N}\\ 0 & , \; elsewhere\end{cases} , \qquad (A.20)$$

satisfying the normalization condition of probability:

$$\int_{-\infty}^{\infty} p_x(x)dx = \int_{-1/\sqrt{2}N}^{1/\sqrt{2}N} \frac{N}{\sqrt{2}}dx = \frac{N}{\sqrt{2}} \int_{-1/\sqrt{2}N}^{1/\sqrt{2}N} dx = \frac{N}{\sqrt{2}} 2 \int_{0}^{1/\sqrt{2}N} dx = \frac{N}{\sqrt{2}} 2 \frac{1}{\sqrt{2}N} = 1. \quad \text{(A.21)}$$

Assume that the imaginary and real parts, $x_I[n]$ and $x_R[n]$, of the complex sequence $x[n]$ (A.12) are two statistical independent variables, so that their joint probability density

$$p_{x_I,x_R}(x_I,x_R) = p_{x_I}(x_I) \cdot p_{x_R}(x_R) \quad \Rightarrow \quad \int_{-\infty}^{\infty}\int_{-\infty}^{\infty} p_{x_I,x_R}(x_I,x_R)dx_I dx_R = 1 \quad \text{(A.22)}$$

can be reduced to the product of the marginal densities:

$$p_{x_I}(x_I) = \int_{-\infty}^{\infty} p_{x_I,x_R}(x_I,x_R)dx_R \quad \Rightarrow \quad \int_{-\infty}^{\infty} p_{x_I}(x_I)dx_I = 1. \quad \text{(A.23)}$$

$$p_{x_R}(x_R) = \int_{-\infty}^{\infty} p_{x_I,x_R}(x_I,x_R)dx_I \quad \Rightarrow \quad \int_{-\infty}^{\infty} p_{x_R}(x_R)dx_R = 1, \quad \text{(A.24)}$$

Applying the probability theory [Papoulis, 1991]:

1. If the joint density of probability for the sequence imaginary and real parts, $x_I[n]$ and $x_R[n]$, is $p_{x_I,x_R}(x_I,x_R)$ (A.22), then the probability density of their ratio (linked to the sequence phase) $y[n]=x_I[n]/x_R[n](=tg\varphi[n])$ can be calculated as

$$p_y(y) = \int_{-\infty}^{\infty} |x_R|\, p_{x_I,x_R}(yx_R,x_R)dx_R \overset{eq.(A.22)}{=} \int_{-\infty}^{\infty} |x_R|\, p_{x_I}(yx_R)p_{x_R}(x_R)dx_R =$$

$$\overset{eq.(A.20)}{=} \int_{-\infty}^{\infty} |x_R| \frac{N}{\sqrt{2}} rect_{1/\sqrt{2}N}(yx_R) \frac{N}{\sqrt{2}} rect_{1/\sqrt{2}N}(x_R)dx_R = \frac{N^2}{2} \int_{-1/(\sqrt{2}Ny)}^{1/(\sqrt{2}Ny)} |x_R|\, rect_{1/\sqrt{2}N}(x_R)dx_R = \quad ,$$

$$= \begin{cases} \dfrac{N^2}{2} \displaystyle\int_{-1/(\sqrt{2}N)}^{1/(\sqrt{2}N)} |x_R|\, dx_R = N^2 \int_{0}^{1/(\sqrt{2}N)} x_R dx_R = \frac{1}{4} & , \quad |y| < 1 \\ \dfrac{N^2}{2} \displaystyle\int_{-1/(\sqrt{2}Ny)}^{1/(\sqrt{2}Ny)} |x_R|\, dx_R = N^2 \int_{0}^{1/(\sqrt{2}Ny)} x_R dx_R = \frac{1}{4y^2} & , \quad |y| > 1 \end{cases}$$

(A.25)

satisfying the probability normalization condition:

$$\int_{-\infty}^{\infty} p_y(y)dy = \int_{-1}^{1} \frac{1}{4}dy + \int_{-\infty}^{-1} \frac{1}{4y^2}dy + \int_{1}^{\infty} \frac{1}{4y^2}dy = 1. \quad \text{(A.26)}$$

2. Introduce the strictly monotonic increasing function $\varphi=f(y)=arctg(y)$, with continuous first derivative $f'(y)=1/(1+y^2)$. If the density of probability for the ratio $y$ is $p_y(y)$ (A.25), then the probability density of the sequence phase $\varphi[n]=arctg(y[n])$ can be calculated as:

$$p_\varphi(\varphi) = \left.\frac{p_y(y)}{f'(y)}\right|_{y=f^{-1}(\varphi)} = p_y(y)(1+y^2)\Big|_{y=tg\varphi} = p_y(tg\varphi)(1+tg^2\varphi) =$$

$$\overset{eq.(A.25)}{=} \begin{cases} \dfrac{1}{4}(1+tg^2\varphi) & , \quad |tg\varphi| < 1 \;\Leftrightarrow\; 0 < \varphi < \dfrac{\pi}{4} \text{ and } \pi - \dfrac{\pi}{4} < \varphi < \pi \\ \dfrac{1}{4tg^2\varphi}(1+tg^2\varphi) = \dfrac{1}{4}(1+\dfrac{1}{tg^2\varphi}) & , \quad |tg\varphi| > 1 \;\Leftrightarrow\; \dfrac{\pi}{4} < \varphi < \pi - \dfrac{\pi}{4} \end{cases} ,$$

(A.27)

satisfying the normalization condition:

$$\int_{0}^{\pi} p_\varphi(\varphi)d\varphi = \int_{0}^{\pi/4} \frac{1}{4}(1+tg^2\varphi)d\varphi + \int_{\pi-(\pi/4)}^{\pi} \frac{1}{4}(1+tg^2\varphi)d\varphi + \int_{\pi/4}^{\pi-(\pi/4)} \frac{1}{4}(1+\frac{1}{tg^2\varphi})d\varphi = 1. \quad \text{(A.28)}$$

It follows that the statistical distribution for the phase $\varphi[n]$ of the stochastic complex sequence $x[n]$ results characterized by the mean value

$$\left\langle \varphi[n]\right\rangle = \int_0^{\pi} \varphi p_{\varphi}(\varphi) d\varphi =$$
$$= \int_0^{\pi/4} \frac{\varphi}{4}(1+tg^2\varphi)d\varphi + \int_{\pi-(\pi/4)}^{\pi} \frac{\varphi}{4}(1+tg^2\varphi)d\varphi + \int_{\pi/4}^{\pi-(\pi/4)} \frac{\varphi}{4}(1+\frac{1}{tg^2\varphi})d\varphi = \quad , \tag{A.29}$$
$$= \frac{1}{16}(\pi - 2\ln 2) + \frac{1}{16}(3\pi + 2\ln 2) + \frac{\pi}{4} = \frac{\pi}{2}$$

then by the mean square value

$$\left\langle \varphi^2[n]\right\rangle = \int_0^{\pi} \varphi^2 p_{\varphi}(\varphi) d\varphi =$$
$$= \int_0^{\pi/4} \frac{\varphi^2}{4}(1+tg^2\varphi)d\varphi + \int_{\pi-(\pi/4)}^{\pi} \frac{\varphi^2}{4}(1+tg^2\varphi)d\varphi + \int_{\pi/4}^{\pi-(\pi/4)} \frac{\varphi^2}{4}(1+\frac{1}{tg^2\varphi})d\varphi = \quad , \tag{A.30}$$
$$= -C + \frac{\pi}{16}(5\pi + \ln 256) \cong \pi$$

being $C$ the Catalan's constant:

$$C = \sum_{m=0}^{\infty} \frac{(-1)^m}{(2m+1)^2} \cong 1 . \tag{A.31}$$

Finally, by the variance:

$$\sigma_{\varphi}^2 = \left\langle \varphi^2[n]\right\rangle - \left\langle \varphi[n]\right\rangle^2 \cong \pi - \frac{\pi^2}{4} . \tag{A.32}$$

Thirdly, consider an index $k$ which assumes $N$ values [see eq. (A.10)],

$$k = 0,1,\dots N-1 , \tag{A.33}$$

with uniform probability:

$$p_k = \frac{1}{N} . \tag{A.34}$$

Similarly to what has been demonstrated by Gauss, the $N$ integer numbers (A.33) satisfy the property for which their sum can be expressed as

$$\sum_{k=0}^{N-1} k = \frac{N(N-1)}{2} , \tag{A.35}$$

and the sum of their squares in the explicit formula:

$$\sum_{k=0}^{N-1} k^2 = \frac{N(N-1)(2N-1)}{6} . \tag{A.36}$$

It follows that the statistical distribution of the $N$ integer numbers (A.33)-(A.34) is characterized by the mean value

$$\left\langle k \right\rangle = \sum_{k=0}^{N-1} k p_k = \frac{1}{N}\sum_{k=0}^{N-1} k \overset{eq.(A.25)}{=} \frac{1}{N}\frac{N(N-1)}{2} = \frac{N-1}{2} , \tag{A.37}$$

then by the mean square value

$$\left\langle k^2 \right\rangle = \sum_{k=0}^{N-1} k^2 p_k = \frac{1}{N}\sum_{k=0}^{N-1} k^2 \overset{eq.(A.36)}{=} \frac{1}{N}\frac{N(N-1)(2N-1)}{6} = \frac{(N-1)(2N-1)}{6} , \tag{A.38}$$

and finally by the variance:

$$\sigma_k^2 = \left\langle k^2 \right\rangle - \left\langle k \right\rangle^2 \overset{eqs.(A.37)-(A.38)}{=} \frac{(N-1)(2N-1)}{6} - (\frac{N-1}{2})^2 = \frac{N^2-1}{12} . \tag{A.39}$$

Concluding, for large values of $N$,

$$\left\langle \Phi[k] \right\rangle \overset{eq.(A.17)}{\cong} \frac{1}{N}\left\{ \sum_{n=0}^{N-1} \left\langle \varphi[n] \right\rangle - \pi \left\langle k \right\rangle (N-1) \right\} =$$
$$\overset{eq.(A.29)}{=} \frac{1}{N}\left\{ \sum_{n=0}^{N-1} \frac{\pi}{2} - \pi \left\langle k \right\rangle (N-1) \right\} = \frac{1}{N}[N\frac{\pi}{2} - \pi \left\langle k \right\rangle (N-1)] =$$
$$\overset{eq.(A.37)}{=} \frac{1}{N}[N\frac{\pi}{2} - \pi \frac{N-1}{2}(N-1)] = \frac{\pi}{2}[1 - \frac{(N-1)^2}{N}] \approx$$
$$\overset{N>>1}{\approx} \frac{\pi}{2}(1-N) \approx -\frac{\pi}{2}N \quad , \qquad \text{(A.40)}$$

$$\sigma_{\Phi}^2 \overset{eq.(A.18)}{\cong} \frac{1}{N^2}[N\sigma_{\varphi}^2 + \pi^2\sigma_k^2 (N-1)^2] \cong$$
$$\overset{eq.(A.32)}{\cong} \frac{1}{N^2}[N(\pi - \frac{\pi^2}{4}) + \pi^2\sigma_k^2 (N-1)^2] =$$
$$\overset{eq.(A.39)}{=} \frac{1}{N^2}[N(\pi - \frac{\pi^2}{4}) + \pi^2 \frac{N^2-1}{12}(N-1)^2] \approx$$
$$\overset{N>>1}{\approx} \frac{\pi^2}{12}\frac{(N^2-1)(N-1)^2}{N^2} \approx \frac{\pi^2}{12}N^2 \quad , \qquad \text{(A.41)}$$

$$\left\langle \Phi^2[k] \right\rangle \overset{eq.(A.19)}{=} \sigma_{\Phi}^2 + \left\langle \Phi[k] \right\rangle^2 \overset{eqs.(A.40)-(A.41)}{\approx} \frac{\pi^2}{12}N^2 + \frac{\pi^2}{4}N^2 = \frac{\pi^2}{3}N^2 , \qquad \text{(A.42)}$$

this appendix has demonstrated a simple formula expressing the phase noise-to-signal ratio due a round-off of FFT processor (the first time on INGV scientific publications, to the best of author knowledge)[see eq. (3.2)]:

$$\frac{\left\langle \left| F[k] \right|^2 \right\rangle}{\left\langle \left| \Phi[k] \right|^2 \right\rangle} \overset{eq.(A.1)-(A.3),(A.42)}{\approx} \frac{N\frac{1}{3}\cdot 2^{-2n_{FFT}}}{\frac{\pi^2}{3}N^2} = \frac{1}{\pi^2 N 2^{2n_{FFT}}} . \qquad \text{(A.43)}$$

**Acknowledgments.**

Dr. A. Settimi would like to thank Drs. C. Bianchi, A. Zirizzotti for the interesting discussions on RESPER probe and Dr.  J. A. Baskaradas for the useful hints on literature acculturating about FFT algorithms.

## References.

Arpaia, P., Daponte, P. and Michaeli, L., (1999). Influence of the architecture on ADC error modelling. IEEE T. Instrum. Meas, 48, 956-966.

Arpaia, P., Daponte, P. and Rapuano, S., (2003). A state of the art on ADC modelling. Comput. Stand. Int., 26, 31–42.

Björsell, N. and Händel, P., (2008). Achievable ADC performance by post-correction utilizing dynamic modeling of the integral nonlinearity. Eurasip J. Adv. Sig. Pr., 2008, ID 497187 (10 pp).

Declerk, P., (1995). Bibliographic study of georadar principles, applications, advantages, and inconvenience. NDT & E Int., 28, 390-442 (in French, English abstract).

Del Vento, D. and Vannaroni, G., (2005). Evaluation of a mutual impedance probe to search for water ice in the Martian shallow subsoil. Rev. Sci. Instrum., 76, 084504 (1-8).

Dishan, H., (1995). Phase Error in Fast Fourier Transform Analysis. Mech. Syst. Signal Pr., 9, 113-118.

Grard, R., (1990). A quadrupolar array for measuring the complex permittivity of the ground: application to earth prospection and planetary exploration. Meas. Sci. Technol., 1, 295-301.

Grard, R., (1990). A quadrupole system for measuring in situ the complex permittvity of materials: application to penetrators and landers for planetary exploration. Meas. Sci. Technol., 1, 801-806.

Grard, R. and Tabbagh, A., (1991). A mobile four electrode array and its application to the electrical survey of planetary grounds at shallow depth. J. Geophys. Res., 96, 4117-4123.

Jankovic, D. and Öhman, J., (2001). Extraction of in-phase and quadrature components by IF-sampling. Department of Signals and Systems, Cahlmers University of Technology, Goteborg (carried out at Ericson Microwave System AB).

Kuffel, J., Malewsky, R. and Van Heeswijk, R. G., (1991). Modelling of the dynamic performance of transient recorders used for high voltage impulse tests. IEEE T. Power Deliver., 6, 507-515.

Ming, X. and Kang, D., (1996). Corrections for frequency, amplitude and phase in Fast Fourier transform of harmonic signal. Mech. Syst. Signal Pr., 10, 211-221.

Mojid, M. A., Wyseure, G. C. L. and Rose, D. A., (2003). Electrical conductivity problems associated with time-domain reflectometry (TDR) measurement in geotechnical engineering. Geotech. Geo. Eng., 21, 243-258.

Mojid, M. A. and Cho, H., (2004). Evaluation of the time-domain reflectometry (TDR)-measured composite dielectric constant of root-mixed soils for estimating soil-water content and root density. J. Hydrol., 295, 263–275.

Oppenheim, A. V., Schafer, R.W. and Buck, J. R., (1999). Discrete-Time Signal Processing (Prentice Hall International, Inc., New York - II Ed.).

Papoulis, A., (1991). Probability, Random Variables, and Stochastic Processes (McGraw-Hill International Editors, Singapore - III Ed.).

Polge, R. J., Bhagavan, B. K. and Callas, L., (1975). Evaluating analog-to-digital converters. Simulation, 24, 81-86.

Razavi, B., (1995). Principles of Data Conversion System Design (IEEE Press, New York).

Samouëlian, A., Cousin, I., Tabbagh, A., Bruand, A. and Richard, G., (2005). Electrical resistivity survey in soil science: a review. Soil Till,. Res., 83, 172-193.

Sbartaï, Z. M., Laurens, S., Balayssac, J. P., Arliguie, G. and Ballivy, G., (2006). Ability of the direct wave of radar ground-coupled antenna for NDT of concrete structures. NDT & E Int., 39, 400-407.

Settimi, A., Zirizzotti A., Baskaradas, J. A. and Bianchi, C., (2010). Inaccuracy assessment for simultaneous measurement of resistivity and permittivity applying sensitivity and transfer function approaches. Ann. Geophys-Italy, 53, 2, 1-19; ibid., Earth-prints, http://hdl.handle.net/2122/5180 (2009); ibid., arXiv:0908.0641 [physics.geophysiscs] (2009).

Settimi, A., Zirizzotti A., Baskaradas, J. A. and Bianchi, C., (2010). Optimal requirements of a data acquisition system for a quadrupolar probe employed in electrical spectroscopy, accepted for publication on Ann. Geophys- -Italy (23/07/2010); ibid, Earth-prints, http://hdl.handle.net/2122/5176 (2009); ibid., arXiv:0908.0648 [physics.geophysiscs] (2009).

Settimi, A., Zirizzotti A., Baskaradas, J. A. and Bianchi, C., (2010). Design of an induction probe for simultaneous measurements of permittivity and resistivity. Quaderni di Geofisica, 79, 26 pp; ibid., Earth-prints, http://hdl.handle.net/2122/5173 (2009); ibid., arXiv:0908.0651 [physics.geophysiscs] (2009).

Tabbagh., A., Hesse, A. and Grard, R., (1993). Determination of electrical properties of the ground at shallow depth with an electrostatic quadrupole: field trials on archaeological sites. Geophys. Prospect., 41, 579-597.

Vannaroni, G., Pettinelli, E., Ottonello, C., Cereti, A., Della Monica, G., Del Vento, D., Di Lellis, A. M., Di Maio, R., Filippini, R., Galli, A., Menghini, A., Orosei, R., Orsini, S., Pagnan, S., Paolucci, F., Pisani, A. R., Schettini, G., Storini, M. and Tacconi, G., (2004). MUSES: multi-sensor soil electromagnetic sounding. Planet. Space Sci., 52, 67–78.

Zhang, J. Q. and Ovaska, S. J., (1998). ADC characterization by an eigenvalues method. Instrumentation and Measurement Technology Conference (IEEE), 2, 1198-1202.